\documentclass[11pt]{amsart}
\usepackage{amsthm}

\usepackage[all]{xy}
\usepackage{amssymb}
\usepackage{enumerate}
\usepackage{mathrsfs}
\usepackage{epsfig}

\usepackage{graphics}
\usepackage{graphicx}
\usepackage{float}
\usepackage{caption}
\usepackage{subcaption}
\usepackage{epigraph}

\numberwithin{equation}{section}

\newcommand{\eq}[1]{(\ref{#1})}

\renewcommand{\Re}{\operatorname{\rm Re}}

\newcommand{\beqast}{\begin{eqnarray*}}
\newcommand{\eqast}{\end{eqnarray*}}
\newcommand{\beqa}{\begin{eqnarray}}
\newcommand{\eqa}{\end{eqnarray}}

\newcommand{\bbe}{\begin{equation}}
\newcommand{\ee}{\end{equation}}

\renewcommand{\Re}{\operatorname{\rm Re}}

\newcommand{\bC}{{\mathbb C}}

\newcommand{\bH}{{\mathbb H}}

\newcommand{\bR}{{\mathbb R}}

\newcommand{\cH}{{\mathcal H}}

\newcommand{\cE}{{\mathcal E}}

\newcommand{\tV}{{\tilde V}}

\newcommand{\la}{\lambda}

\newcommand{\La}{\Lambda}

\newcommand{\sg}{\sigma}

\newcommand{\ga}{\gamma}

\newcommand{\dd}{\partial}

\newcommand{\bfo}{{\bf 1}}

\begin{document}

\title[Alternative models for FX and double barrier options in L\'evy models]
{Alternative models for FX: pricing  double barrier options in regime-switching L\'evy models with memory}
\author[
Svetlana Boyarchenko and
Sergei Levendorski\u{i}]
{
Svetlana Boyarchenko and
Sergei Levendorski\u{i}}

\begin{abstract}
This paper is a supplement to our recent  paper ``Alternative models for FX, arbitrage opportunities  and efficient pricing of double barrier options in L\'evy models". 
We introduce the class of regime-switching L\'evy models with memory, which take into account the evolution of the stochastic parameters
in the past. This generalization of the class of L\'evy models modulated by Markov chains is similar in spirit to rough volatility models. It is flexible  and suitable for application of the machine-learning tools. We formulate the modification of the numerical method in ``Alternative models for FX, arbitrage opportunities  and efficient pricing of double barrier options in L\'evy models", which has the same number of the main time-consuming blocks as  the method for Markovian regime-switching models.

\end{abstract}

\thanks{
\emph{S.B.:} Department of Economics, The
University of Texas at Austin, 2225 Speedway Stop C3100, Austin,
TX 78712--0301, {\tt sboyarch@utexas.edu} \\
\emph{S.L.:}
Calico Science Consulting. Austin, TX.
 Email address: {\tt
levendorskii@gmail.com}}

\maketitle

\noindent
{\sc Key words:} regime-switching L\'evy processes,  double barrier options, Wiener-Hopf factorization, Fourier transform, Laplace transform, 
 Gaver-Wynn Rho algorithm, sinh-acceleration

\noindent
{\sc MSC2020 codes:} 60-08,42A38,42B10,44A10,65R10,65G51,91G20,91G60


\section{Introduction}\label{s:intro}
 In \cite{AltFX}, an efficient numerical method
for pricing double-barrier options in regime-switching L\'evy models is developed.  
This text can be regarded as an additional section in \cite{AltFX}. We consider a natural generalization
of the regime-switching L\'evy models: the evolution of the parameters of the L\'evy model depends on the realizations of stochastic parameters in the past. This is similar in spirit to 
rough volatility models \cite{RoughVolBook}, and one should expect that approximations of the latter
by regime-switching models with memory is possible. Flexibility of regime-switching models with memory makes this type of models a good candidate for application of machine learning tools \cite{MachineLearnFinance}.

After the truncation of histories, which is essentially unavoidable for a numerical realization, regime-switching models with memory constitute a subclass of standard regime-switching models.
However, at each step of the iteration procedure  in \cite{AltFX}, it is necessary to calculate the price of  double-barrier options for each state of the modulating Markov chain. This requires the evaluation of the Wiener-Hopf factors
and numerical realization of operators for each state. These main blocks are time-consuming, hence, if they are different for each history, the CPU time becomes extremely large or parallelization complicated. We use the modification which uses the blocks depending on
the current realization of stochastic factors but not on the realizations in the past. The other operations in the iteration procedure are evaluation of scalar products
of precalculated arrays of transition rates and value functions calculated at the previous step 
 of the iteration procedure. As the final step - the inverse Laplace-Fourier transform of vector-functions - these operations are  easily parallelisable.
 
 In order not to copy-paste necessary preliminary pieces from  \cite{AltFX}, we recall the main structure 
 of our approach \cite{Contrarian,EfficientLevyExtremum,EfficientDoubleBarrier,AltFX} to pricing barrier options.
 Applying the Laplace transform, equivalently, randomizing the maturity date, we reduce the problem of pricing a barrier option (with a single barrier or two barriers) to evaluation of the corresponding perpetual barrier option in the L\'evy model
 or regime-switching model; the Laplace transform $\tV(q)$ of the option price admits analytic continuation w.r.t. the spectral parameter $q$ to the right-half plane, and, in the case of sufficiently regular L\'evy processes, to a sector $\Sigma_\ga+\sg_0$,
 where
 $\Sigma_\ga=\{z=\rho e^{i\varphi}|\ |\varphi|<\ga, \rho>0\}$, and $\sg_0>0, \ga>\pi/2$. 
  Assuming that $\tV(q)$ can be efficiently evaluated for each $q$ used in the chosen Laplace inversion algorithm, the most efficient algorithm is based 
 on the conformal deformation of the contour in the Bromwich integral (sinh-acceleration). The deformation is possible 
 if $\tV(q)$ admits analytic continuation to $\Sigma_\ga+\sg_0$.  If $\tV(q)$ does not admit analytic continuation a sector of the form $\Sigma_\ga+\sg_0$, we apply the GWR-algorithm (Gaver-Stehfest algorithm with the Wynn-Rho acceleration).
 
For each $q$, we calculate $\tV(q)$  using an iteration procedure  (in the case of regime-switching models, each step is an additional iteration procedure).
The main blocks for the evaluation of $\tV(q)$ are
calculation of the Wiener-Hopf factors for each L\'evy process,
and evaluation of first touch digital options in each L\'evy model using the EPV-operators technique. In the case of regime-switching models,
an additional element of the algorithm are
evaluation of scalar products of vectors of transition rates and vectors of value functions calculated 
at the preliminary step.  
Efficient calculations are possible if the L\'evy processes are SINH-regular; in the case of a subclass of regular Stieltjes-L\'evy processes,
(SL-processes) calculations are more efficient.
 See \cite{SINHregular,EfficientAmenable} for the definitions. In \cite{EfficientAmenable},
 it is shown that essentially all popular classes of L\'evy processes bar stable L\'evy processes are regular SL-processes.
The deformation of the contour of integration in the Bromwich integral is possible if
SINH-regular processes are of infinite variation or drfitless processes of finite variation. The algorithms
in \cite{EfficientLevyExtremum,EfficientDoubleBarrier} are designed for these processes. The method
in \cite{AltFX} uses the GWR algorithm and the details are spelled out for processes of finite variation with non-zero drift.
However, the  evaluation of $\tV(q)$ in \cite{AltFX} for regime-switching models can be used when the sinh-deformation of the contour of integration in the Bromwich integral is possible. The same remark is valid for the extension
to the regime-switching L\'evy processes with memory in this paper.

 \section{Evaluation of the perpetual double barrier options}
We generalize the setting of \cite{AltFX} allowing for the transition rates to depend on the history -
the sequence of states $h=(h_0,h_{-1},\ldots), h_j\in \{1,2,\ldots, m\}$ visited by the process $Y$ in the past\footnote{We apologies to the reader for using the same letter $h$ to denote
the barriers and histories. The barriers have the subscripts $\pm$, the history have subscripts $0,-1,\ldots $ only.}. The
main restriction on the history is $h_{-j}\neq h_{-j-1}, j=0,-1,\ldots$. We can regard $Y$ as the process on 
the countable state space $\bH$ of strategies, the transition rates  from state $h^j$ to state
$h^k$ being zero unless $h^k_{-\ell}=h^j_{1-\ell}, \ell=-1,-2,\ldots$. Hence, it suffices to
introduce the notation $\la_{s,h}$ for the transition rate from $h$ to $(s, h_0,h_{-1},\ldots)$, where $h\in \bH$ and 
$s\neq h_0$. 
Since a numerical realization is possible only after an
appropriate truncation of histories, we assume that the process ``remembers" only the last $N$ states visited.
The histories are of the form $(h_0,h_{-1},\ldots, h_{-N})$.
The set of histories is denoted $\bH_N$. We have $\# \bH_N=m\cdot (m-1)^N$.  For $h\in\bH_N$ and $s\in \{1,2,\ldots,m\}$, denote
$h'=(h_0,h_{-1},\ldots, h_{-N+1})$ and $(s,h')=(s,h_0,h_{-1},\ldots, h_{-N+1})\in \bH_N$.
If the process $Y$ approximates a diffusion, then an additional natural restriction is $\la_{s,h}=0$ unless $s\in \{h_0-1,h_0+1\}$, and $\# \bH_N<m\cdot 2^N$. A very large $\# \bH_N$ is less of a problem as one would expect because
the main block of the method is the evaluation of the perpetual double-barrier options in the L\'evy models
with the infinitesimal generators $L_j$, $j=1,2,\ldots, m$, which admits a straightforward parallelization. 
We slightly modify the construction in \cite{AltFX} as follows. 

Denote by $V_h(t,x)$ the value function at time $t$ and $X_t=x$, after the history $h\in\bH_N$, and set 
$\La_{h}=\sum_{s\neq h(0)}\la_{s,h}$, $Q_h(q)=q+\La_h+r_{h_0}$. 
The vector-function $\{V_h(t,x)\}_{h\in\bH_N}$, $t<T$, is the solution of the system
\beqa\label{rswM1}
(\dd_t+L_{h_0}-r_{h_0}-\La_{h})V_h(t,x)&=&-\sum_{s\neq h(0)}\la_{s,h}V_{(s,h')}(t,x),\ t<T, x\in (h_-,h_+),\\\label{rswM2}
V_h(T,x)&=&G_{h_0}, \hskip3cm x\in (h_-,h_+),\\\label{rswM3}
V_h(t,x)&=&0, \hskip2.5cm t\le T,\ x\not\in (h_-,h_+).
\eqa 
Applying the Laplace transform w.r.t. $\tau=T-t$,
we obtain the system 
\beqa\label{rsqwM12}
(Q_h(q)-L_{h_0})\tV_h(q,x)&=&G_{h(0)}+\sum_{s\neq h(0)}\la_{s,h}\tV_{(s,h')}(q,x),\ x\in (h_-,h_+),
\\\label{rsqwM32}
\tV_h(q,x)&=&0, \hskip2.5cm  x\not\in (h_-,h_+).
\eqa
For $q>0$, denote by $\tV^0(q)\in \bR^{\#\bH_N}$ the unique solution of the system
\bbe\label{MMV0j}
Q_h(q)\tV^{0}_h(q)= G_{h_0}+\sum_{s\neq h(0)}\la_{s,h} \tV^{0}_{(s,h')}(q),\ h\in \bH_N,
\ee
and set $\tV^{1}_h(q,x)=\tV_h(q,x)-V^{0}_h(q)$. The vector-function $\tV^{1}(q,\cdot)=[\tV^{1}_h(q,\cdot)]_{h\in\bH_N}$
is a unique bounded solution of the system
\beqa\label{rsqwM13}
(Q_h(q)-L_{h_0})\tV^{1}_h(q,x)&=&\sum_{s\neq h(0)}\la_{s,h}\tV^{1}_{(s,h')}(q,x), \quad x\in (h_-,h_+),
\\\label{rsqwM33}
\tV^{1}_h(q,x)&=&-\tV^{0}_h(q), \hskip1.5cm \ x\not\in (h_-,h_+).
\eqa 
The system \eq{rsqwM13}-\eq{rsqwM33} can be solved  
as a similar system in \cite{AltFX} for regime-switchning models. However, in this case, at each step of the iteration procedure,
it is necessary to solve $\# \bH_N$ problems, with the operators
$L_{h(0)}-r_{h_0}-\La_h$ depending on $h\in\bH_N$. Since  $L_{h(0)}$ depends on the current realization of the modulating process only, the number of operators, in particular, the number
of the Wiener-Hopf factorization blocks, can be greatly decreased.
Let  there exist $\La_0>0$ such that
\bbe\label{boundLa0}
|\La_{s,h}|\le \La_0,\  h\in\cH, s\neq h_0.
\ee
We set $Q(s; q):=q+\La_0+r_{s},$ $s=1,2,\ldots, m$, 
and rewrite \eq{rsqwM13} as follows: for $h\in \bH_N$,
\bbe\label{rsqwM14}
(Q(h_0; q)-L_{h_0})\tV^{1}_h(q,x)=(\La_0-\La_h)\tV^{1}_h(q,x)+\sum_{s\neq h(0)}\la_{s,h}\tV^{1}_{(s,h')}(q,x),
\  x\in (h_-,h_+).
\ee
A new term on the RHS appears but the operators on the LHS depend on $h_0$ only, hence, there are only $m$
different operators.
We calculate $\tV^1=[\tV^1_h]_{h\in\bH_N}$ in the form of the series 
\beqa\label{qtVrsw}
 \tV^1(q,x)&=&\sum_{\ell =1}^{+\infty} (-1)^\ell (\tV^{+;\ell}(q,x)+\tV^{-;\ell}(q,x)).
 \eqa
 Set $\tV^{-;0}(q,x)=\bfo_{[h_+,+\infty)}(x)\tV^0(q)$ and 
$\tV^{+;0}(q,x)=\bfo_{(-\infty,h_-]}(x)\tV^0(q)$. For $\ell=1,2,\ldots,$ inductively define
$\tV^{\pm;\ell}(q,x)=[\tV^{\pm;\ell}_{h}(q,x)]_{h\in\bH_N}$ as the unique bounded
solution of the system
\beqa\label{rsqwp1M}
(Q(h_0; q)-L_{h_0})\tV^{+;\ell}_{h}(q,x)&=&(\La_0-\La_h)\tV^{+;\ell}_h(q,x)\\\nonumber &&+\sum_{s\neq h(0)}\la_{s,h}\tV^{+;\ell}_{(s,h')}(q,x), \ x<h_+,
\\\label{rsqwp2M}
\tV^{+;\ell}_h(q,x)&=&\tV^{-;\ell-1}_h(q,x), \hskip0.8cm \ x\ge h_+,
\eqa
and
\beqa\label{rsqwm1M}
(Q(h_0; q)-L_{h_0})\tV^{-;\ell}_{h}(q,x)&=&(\La_0-\La_h)\tV^{-;\ell}_h(q,x)\\\nonumber &&+\sum_{s\neq h(0)}\la_{s,h}\tV^{-;\ell}_{(s,h')}(q,x), \ x>h_-,
\\\label{rsqwm2M}
\tV^{-;\ell}_h(q,x)&=&\tV^{+;\ell-1}_h(q,x), \hskip0.8cm \ x\le h_-.
\eqa
Let $\cE_{Q(h_0; q)}, \cE^\pm_{Q(h_0;q)}$ be 
the EPV operators under $X^{h_0}$, the discount rate being $Q(h_0;q)$. The general theorems for single barrier options in \cite{barrier-RLPE,NG-MBS}
(see also  \cite[Thm's 11.4.2-11.4.5]{IDUU})
allow us to rewrite the boundary problems \eq{rsqwp1M}-\eq{rsqwp2M} and \eq{rsqwm1M}-\eq{rsqwm2M} in the form
\beqa\label{rsqwp12M}
\tV^{+;\ell}_{j}(q,x)&=&\frac{1}{Q(h_0;q)}\cE^+_{Q(h_0;q)}\bfo_{(-\infty,h_+)}\cE^-_{Q(h_0;q)}
\\\nonumber
&&\left((\La_0-\La_h)\tV^{+;\ell}_h(q,x)+\sum_{s\neq h_0}\la_{s,h}\tV^{+;\ell}_{(s,h')}(q,x)\right)
\\\nonumber
&&+\cE^+_{Q(h_0;q)}\bfo_{[h_+,+\infty)}(\cE^+_{Q(h_0;q)})^{-1}\tV^{-;\ell-1}_h(q,x), \ h\in\bH_N,
 \eqa
 and
 \beqa\label{rsqwm12M}
\tV^{-;\ell}_{j}(q,x)&=&\frac{1}{Q(h_0;q)}\cE^-_{Q(h_0;q)}\bfo_{(h_-,+\infty)}\cE^+_{Q(h_0,q)}
\\\nonumber
&&\left((\La_0-\La_h)\tV^{-;\ell}_h(q,x)+\sum_{s\neq h_0}\la_{s,h}\tV^{-;\ell}_{(s,h')}(q,x)\right)
\\\nonumber
&&+\cE^-_{Q(h_0,q)}\bfo_{(-\infty,h_-]}(\cE^-_{Q(h_0,q)})^{-1}\tV^{+;\ell-1}_h(q,x), \ h\in\bH_N,
 \eqa
 respectively.   The systems \eq{rsqwp12M} and \eq{rsqwm12M} are solved using the straightforward modification
 of the iteration procedure in \cite{AltFX}. Explicitly, let $\ell\ge 1$ be fixed and $\tV^{-;\ell-1}_h(q,\cdot)\in L_\infty(\bR)$, $h\in\bH_N$ be given.  If $\Re q$ is sufficiently large, the RHS' of the system \eq{rsqwp12M} defines a contraction map from $L_\infty(\bR;\bC^{\#\bH_N})$ to $L_\infty(\bR;\bC^{\#\bH_N})$
 (in addition, the map is monotone). Therefore,
 letting $\tV^{+;\ell;0}_{j}(q,\cdot)=0$ and, for $n=1,2,\ldots,$ defining 
 \beqa\label{rsqwp12iterM}
\tV^{+;\ell;n}_{j}(q,x)&=&\frac{1}{Q(h_0;q)}\cE^+_{Q(h_0;q)}\bfo_{(-\infty,h_+)}\cE^-_{Q(h_0;q)}
\\\nonumber
&&\left((\La_0-\La_h)\tV^{+;\ell;n}_h(q,x)+\sum_{s\neq h_0}\la_{s,h}\tV^{+;\ell;n-1}_{(s,h')}(q,x)\right)
\\\nonumber
&&+\cE^+_{Q(h_0;q)}\bfo_{[h_+,+\infty)}(\cE^+_{Q(h_0;q)})^{-1}\tV^{-;\ell-1}_h(q,x), \ h\in\bH_N,
 \eqa
 we conclude that $\tV^{+;\ell}_h(q,\cdot)=\lim_{n\to \infty}\tV^{+;\ell;n}_h(q,\cdot)$. The system \eq{rsqwm12M} is solved similarly:
 we set $\tV^{-;\ell;0}_h(q,\cdot)=0$, $h\in\bH_N$, then, for $\ell=1,2,\ldots,$ define
 \beqa\label{rsqwm12iterM}
\tV^{-;\ell;n}_h(q,x)&=&\frac{1}{Q(h_0;q)}\cE^-_{Q(h_0;q)}\bfo_{(h_-,+\infty)}\cE^+_{Q(h_0;q)}
\\\nonumber && \left((\La_0-\La_h)\tV^{-;\ell;n-1}_h(q,x)+ \sum_{s\neq h(0)}\la_{s,h}\tV^{-;\ell;n-1}_{(s,h')}(q,x)\right)
\\\nonumber
&&+\cE^-_{Q(h_0,q)}\bfo_{(-\infty,h_-]}(\cE^-_{Q(h_0;q)})^{-1}\tV^{+;\ell-1}_{j}(q,x), \ j=1,2,\ldots, m,
 \eqa
 and conclude that $\tV^{-;\ell}_h(q,\cdot)=\lim_{n\to \infty}\tV^{-;\ell;n}_{h}(q,x), h\in\bH_N$.
If the calculations are in the state space, then the same grids can and should be used for the numerical realization
of \eq{rsqwp12iterM} and \eq{rsqwm12iterM}, hence, one can use a straightforward variation of the algorithm  \cite{BLdouble} for the non-regime switching case. The CPU time decreases and accuracy increases if the
calculations are in the dual space as in \cite{AltFX}.

For $s\in \{1,2,\ldots,N\}$, set $\cH_N(s)=\{h\in \bH_N\ |\ h_0=s\}$. At each step of the iteration procedure,
the evaluation of a value function with the subscript $h\in \cH_N(s)$ is easily parallelized - the same operators
are applied to different functions - and the other operations in the iteration procedure are evaluation of scalar products
of precalculated arrays of transition rates and value functions calculated at the previous step 
 of the iteration procedure. The final step - the inverse Laplace-Fourier transform of vector-functions - is also easily parallelizable.


\end{document}